\documentclass[12pt,preprint]{aastex}

\def\lsim{\raise0.3ex\hbox{$<$}\kern-0.75em{\lower0.65ex\hbox{$\sim$}}}
\def\gsim{\raise0.3ex\hbox{$>$}\kern-0.75em{\lower0.65ex\hbox{$\sim$}}}

\begin{document}

\title{Magnetocentrifugal Winds in 3D: Nonaxisymmetric Steady State}
\author{Jeffrey M. Anderson\altaffilmark{1},
Zhi-Yun Li\altaffilmark{1}, Ruben Krasnopolsky\altaffilmark{2},
\& Roger D. Blandford\altaffilmark{3}}
\altaffiltext{1}{Astronomy Department, University of Virginia,
Charlottesville, VA 22904; jma2u, zl4h@virginia.edu}
\altaffiltext{2}{Department of Astronomy \& Astrophysics,
University of Toronto, Toronto, ON M5S 3H4, Canada}
\altaffiltext{3}{SLAC, M/S 75, 2575 Sandhill Rd, Menlo Park, CA 94025;
rdb3@stanford.edu}

\begin{abstract}
Outflows can be loaded and accelerated to high speeds along rapidly
rotating, open magnetic field lines by centrifugal forces. Whether
such magnetocentrifugally driven winds are stable is a longstanding
theoretical problem. As a step towards addressing this problem,
we perform the first large-scale 3D MHD simulations that extend
to a distance $\sim 10^2$ times beyond the launching region,
starting from steady 2D (axisymmetric) solutions. In an attempt
to drive the wind unstable, we increase the mass loading on one
half of the launching surface by a factor of $\sqrt{10}$,
and reduce it by the same factor on the other half. The evolution
of the perturbed wind is followed numerically. We find no evidence
for any rapidly growing instability that could disrupt the wind
during the launching and initial phase of propagation, even when the
magnetic field of the magnetocentrifugal wind is toroidally dominated
all the way to the launching surface. The strongly perturbed wind
settles into a new steady state, with a highly asymmetric mass
distribution. The distribution of magnetic field strength is, in
contrast, much more symmetric. We discuss possible reasons for
the apparent stability, including stabilization by an axial
poloidal magnetic field, which is required to bend field lines
away from the vertical direction and produce a magnetocentrifugal
wind in the first place.

\end{abstract}


\section{Introduction}

The jets and winds observed around young stellar objects (YSOs) are thought
to be driven magnetocentrifugally from disk surface (\citealt{bp82};
see \citealt{us85} for a related mechanism). The fluid
rotation winds the magnetic field up into a predominantly toroidal
configuration at large distances.  The toroidal field is thought to be
able to collimate part of the wind into a jet through ``hoop'' stresses
\citep{shu95,hn89}.
It may, however, lead to instabilities that can
potentially disrupt the outflow \citep{e93,be98}.

How narrow astrophysical jets maintain their stability over large distances
is a longstanding puzzle \citep{f98}.
Numerical simulations have demonstrated that hydrodynamical
jets are prone to disruption by Kelvin-Helmholtz (KH) instabilities (e.g.,
\citealt{bo98,h04}).  Magnetic fields can add rigidity to a flow, and have a
stabilizing effect against KH instabilities.  They may, however, introduce
current-driven (CD) instabilities (e.g., \citealt{ac92,lba00}).
\citet{bk02} studied
the interplay between KH and CD instabilities and concluded that large-scale
deformation of magnetic fields associated with the CD mode can effectively
saturate the KH surface vortices and thus aid in jet survival.
Whether magnetized
jets can indeed travel large distances without being disrupted remains
an area of active research (e.g., \citealt{nm04}).

By comparison, the stability of magnetically driven jets and winds {\it during
launching and early propagation} is less explored.  \citet{lb96} studied the
3D stability of a jet accelerated and pinched by a purely
toroidal magnetic field.  They found that the $m=1$ (kink) instability can
cause the tip of the jet to fold back upon itself.
The mode is stabilized by poloidal magnetic fields in the simulations
of \citet{lb97}, where the jet is squirted along the (initially
uniform) poloidal field lines by the high pressure created near the
central object through rapid equatorial infall. \citet{ocp03} examined
the 3D stability of a cold jet launched magnetically from a Keplerian
disk. They too adopted
an initially uniform magnetic field threading the disk vertically. The
differential rotation between the disk and a stationary (pressure-supported)
corona winds up the field lines, generating a larger toroidal field at
a smaller radius. The gradient
in the toroidal field can bend the initially vertical field lines away from
the disk axis by more than $30\degr$, enabling steady outflow through the
magnetocentrifugal mechanism.  \citet{op99} showed that a relatively large
mass loading is required to generate a large enough toroidal field gradient
to open up the field lines for steady magnetocentrifugal wind driving in
2D; lightly-loaded outflows remain unsteady,
generating knots episodically.  The 3D jet of \citet{ocp03} has parameters
in this episodic regime. More recently, \citet{ks05} carried
out 3D simulations of disk-corona system threaded (again) by vertical
field lines. They found that a jet is produced by the Uchida-Shibata
mechanism, despite the non-axisymmetric perturbations imposed on the
disk rotation rate. In this letter, we are interested in the stability
of cold magnetocentrifugal winds accelerated steadily along field lines
inclined more than $30^\circ$ away from the axis, as in the original
picture of \citet{bp82}.

\section{Simulation Setup and Numerical Results}
\label{desc}

We simulate the disk-driven magnetocentrifugal wind using Cartesian
coordinate system, with the $z$-axis along the rotation axis, and
$x$- and $y$-axis in the disk plane. Our calculations are carried
out using an MPI-parallel version of the ZEUS-3D MHD code \citep{cnf94},
which we have previously used to simulate 2D axisymmetric winds
(\citealt{klb99,klb03a}; \citealt{a05}, Paper I
hereafter). The 2D simulations serve as the starting points for
our new, 3D calculations. They are specified by three functions on
the disk surface: the distributions of disk rotation speed $v_d(\varpi)$,
vertical field strength $B_z(\varpi)$, and rate of mass loading per
unit area $\rho(\varpi) v_z(\varpi)$, where $\varpi$ is the cylindrical
radius from the axis, and $\rho$ and $v_z$ are the density and vertical
component of the injection speed at the base of the wind. Inside a
sphere of radius $r=\varpi_g$, we
soften the gravitational field of the central point mass to avoid
singularity (eq.~[11] of Paper I). The softening yields an
equilibrium disk in  sub-Keplerian rotation inside $\varpi_g$.
Outside $\varpi_g$, the disk rotation is
exactly Keplerian. The magnetocentrifugal wind is
launched from this portion of the disk, from $\varpi_g$ to
an outer radius $\varpi_0$. The material coming off of the
outer edge is assumed to slide outwards along the equatorial
plane to fill all available space.

On the launching surface between $\varpi_g$ and $\varpi_0$, we impose
the Blandford-Payne distributions for density $\rho\propto
\varpi^{-3/2}$ and field strength
$B_z\propto \varpi^{-5/4}$; the latter is multiplied by a spline
function $S(\varpi)$
to bring it to zero at the outer edge of the launching region (see
eq.~[14] of Paper I), as demanded by the space-filling requirement.
Cold material is injected into the wind at a slow speed $v_z=0.1\
V_K (\varpi)\ S(\varpi)$, where $V_K$ is the Keplerian speed.
Inside the softening radius $\varpi_g$,
$B_z$ continues to increase slowly inwards, reaching a maximum value
at the origin. In
this sub-Keplerian region, the field lines are generally not
inclined by a large enough angle away from the axis
to drive a cold outflow centrifugally. Here, we inject a low-density
material along the field lines, with $v_z$ set to twice the local
escape velocity. This tenuous, fast-moving axial flow carries a
small fraction ($8.2\%$) of the total mass flux. It provides a
clean inner boundary to the magnetocentrifugally driven wind, the
focus of our study, and may represent the stellar wind inferred
in some young stellar objects from blue-shifted absorption lines
(e.g., \citealt{e06}).

The 2D wind solutions are characterized by a dimensionless mass-loading
parameter
$
\mu_g= {4\pi \rho v_z V_K/B_z^2},
$
where the density, injection and rotation speeds, and field strength
are evaluated at the radius $\varpi_g$ on the disk. In ``light''
winds with $\mu_g \ll 1$, the magnetic field is initially dominated
by the poloidal component near the launching surface. It becomes
toroidally dominated only outside the Alfv\'en surface \citep{s96}.
As $\mu_g$ approaches unity, the winds become toroidally
dominated all the way to the launching surface. These are ``heavy''
winds. In Paper I, we have explored the structure of 2D winds with
$\mu_g$ varying from $6.25\times 10^{-4}$ to 6.25. Since our main
interest here is to determine whether a 2D wind
is stable in 3D to potential instabilities driven by toroidal field,
we decide to focus on the $\mu_g=0.625$ case, which is representative
of toroidally-dominated heavy winds. The heavy wind has the added
advantage of having a relatively low Alfv\'en speed, which allows for
a relatively large timestep, which in turn enables the simulation to
reach a later physical time than the lighter wind simulations that we
have also performed.

Our simulations are carried out in dimensionless quantities. We set
the inner radius of the Keplerian disk to $\varpi_g=1$, and the outer
radius of the wind-launching region to $\varpi_0=5$. The simulation
box extends
far beyond the launching region, to $\pm 500$ in the $x$ and $y$
directions, and to $400$ in the $z$ direction. We adopt a
$240\times240\times132$ grid, with a uniform sub-grid of
$80\times80\times32$ covering the innermost $12\times12\times4$
region (which includes all of the launching surface), and the
remaining grid spaced logarithmically. As in the 2D
case, we impose conditions on the electromotive force at the
launching surface such that
the field lines are firmly anchored on the disk at their
footpoints while able to twist and bend freely in response to
the stresses in the
wind \citep{klb99,klb03a}. A technique based on Appendix A3
of \citet{ocp03} is used to ensure the anchoring of the field lines
on the disk to machine accuracy in a Cartesian coordinate system.
On the remaining boundaries, the standard outflow conditions in
ZEUS-3D are adopted.

\begin{figure}
\plotone{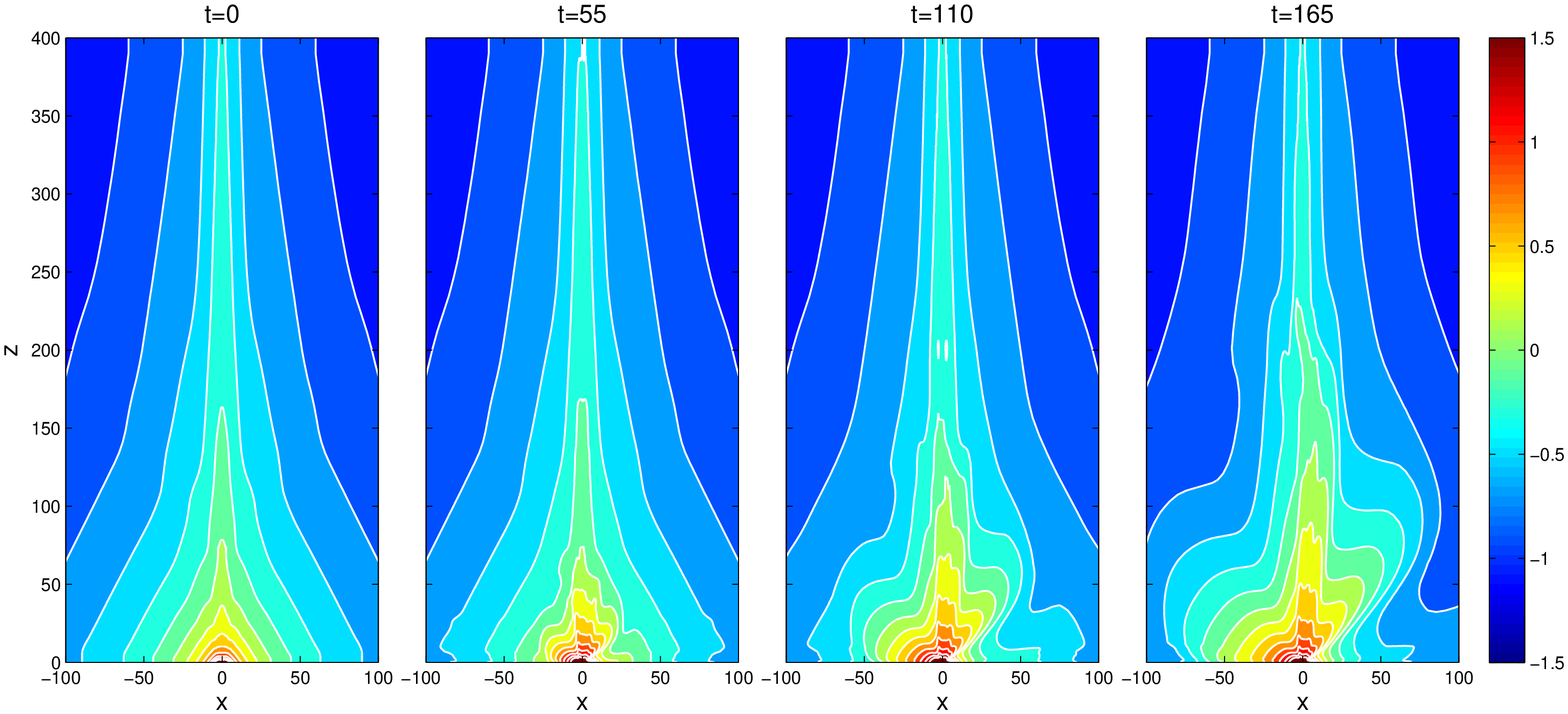}
\caption{Distributions of column density (log scale) in the $y$ direction
at four representative times, with 5 contours per decade. The inner
contours for the last two
times appear nearly identical, signaling the approach to a steady
state.
} \label{fig:f1}
\end{figure}

\begin{figure}
\plotone{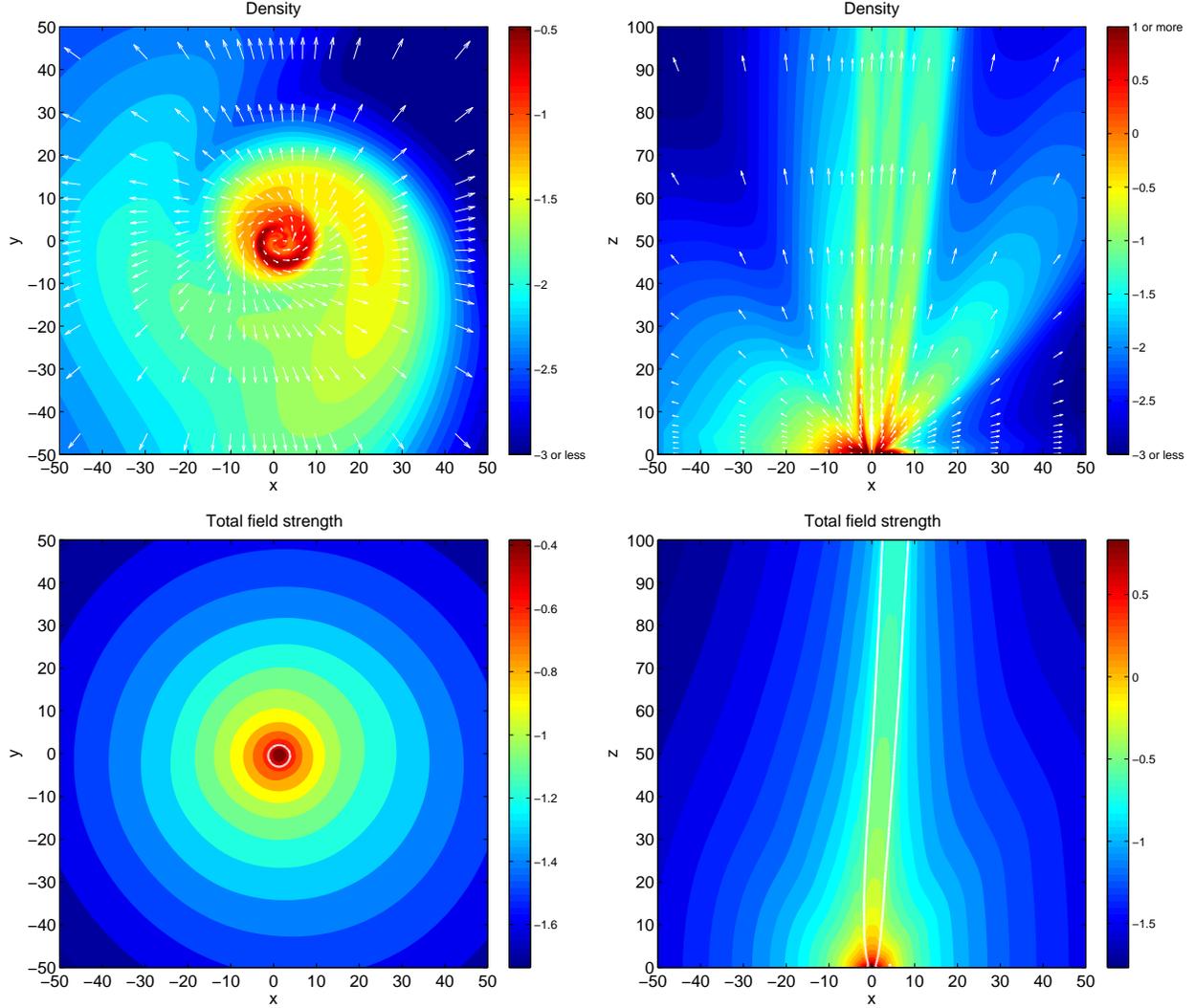}
\caption{Properties of the non-axisymmetric steady state. Upper
panels show the density distribution (log scale) in an $xy$-plane
at height $z=30.14$ (left) and in the $xz$-plane at $y=0$ (right).
Superposed are vectors of velocity field in the plane, with the
length of arrow proportional to the magnitude of velocity.
Lower panels show the distribution of total field strength
(log scale) in the same $xy$- (left) and $xz$-plane (right), with
white contours marking the location where $B_z=(B_x^2+B_y^2)^{1/2}$.
Roughly speaking, the magnetic fields inside (outside) the
contours are poloidally (toroidally) dominated.
} \label{fig:f2}
\end{figure}

We perturb the initially steady axisymmetric wind through the boundary
conditions on the disk between $\varpi_g\le \varpi \le \varpi_0$. We
increase the
mass loading rate (through the density at the base of the wind) by
a factor of $\sqrt{10}$ on one half of the disk ($x \ge 0$) and
decrease it by the same factor on the other half ($x < 0$).
This perturbation is initiated at $t=0$, and is kept throughout
the simulation. The goal is to determine whether the large
non-axisymmetric perturbation imposed at the launching surface
can lead to disruption of the wind during its acceleration and
initial phase of propagation, particularly by the kink ($m=1$) mode.
The numerical results are shown in Figs.~\ref{fig:f1} and \ref{fig:f2}.

Fig.~\ref{fig:f1}
shows the snapshots of column density in the $y$ direction at
four representative times, in units of the radius divided by the
Keplerian speed at $\varpi_g$. (In this unit, the rotation period
at the inner edge of the Keplerian disk is $2\pi$.) At time $t=0$,
the column density in the axisymmetric wind is well collimated at
large distances along the axis, as predicted by \citet{shu95}
from asymptotic analysis. As time progresses, an increasingly large
portion of the wind becomes distorted as the perturbation propagates
outwards from the launching surface. There is no evidence, however,
for any growth of instabilities that are commonly expected for such
a toroidally dominated wind. Indeed, the perturbed region appears to
settle quickly into a {\it nonaxisymmetric} steady state, as can be
seen by comparing the column density contours in the inner parts of
the last two frames.

Fig.~\ref{fig:f2} displays selected properties of the apparent steady state.
In the upper panels, we show the distributions of volume density
in an $xy$-plane at height $z=30$ and in an $xz$-plane at $y=0$
for the time $t=165$, corresponding to $26.3$ times the rotation
period at the inner edge of the Keplerian disk. Clearly, the
density distribution is strongly asymmetric. In the $xy$-plane,
it is dominated by a trailing spiral arm outside the central
region -- the region occupied by the non-magnetocentrifugal
outflow injected near the axis (termed ``the axial column''
hereafter). The spiral is created by the smearing of the denser
material magnetocentrifugally launched from one half of
the disk surface by rotation. Wind rotation is evident from
the velocity vectors displayed, particularly in the central
region where the velocity field is dominated by rotation rather
than outflowing motion. Inside the axial column, there is some
hint of a 4-armed spiral structure in density distribution. We
believe the structure is numerical in origin, since the conditions
at the base of the axial column are kept axisymmetric. Most likely,
it is generated by the rectangular grid, which is not ideal for
following rotating motion near the axis. Nevertheless, the numerical
artifact appears localized inside the axial column. Between the
column and the
surrounding magnetocentrifugal wind lies a shell of high density.
Most likely, it is created by the squeezing of the strong toroidal
magnetic field in the magnetocentrifugal wind against the strong
poloidal field in the column. The compressed shell is
evident in the density distribution in the $xz$-plane.
The shell, which encases the axial column, is
tilted away from the $z$-axis. The dense ridge further to the
right of the axis (at an angle $\sim 45^\circ$) corresponds
to the large-scale spiral arm in the $xy$-plane, which has a
helical shape in 3D.

The strong asymmetry in mass distribution all but disappears in
the distribution of magnetic field, as shown in the lower panels
of Fig.~\ref{fig:f2}. In the $xy$-plane, contours of constant {\it total}
field strength form nearly concentric rings. Close inspection
shows that the rings are shifted slightly in the positive $x$
direction. The white contour near the center marks the location
where $B_z=(B_x^2+B_y^2)^{1/2}$. Roughly speaking, it divides
the axial column of non-magnetocentrifugal outflow (inside the
contour) where the magnetic field is mainly poloidal from the
toroidally dominated magnetocentrifugal part of the wind; the
latter occupies most of the space. The bending of the axial
column can be seen more clearly in the $xz$-plot. Except for the
narrow region inside the white contour, the magnetic field
is toroidally dominated, including the launching surface. The
much more symmetric field distribution indicates that the mechanical
structure of the steady wind is controlled to a large extent
by magnetic stresses, rather than forces due to fluid motions.

\section{Discussion: a Built-in Stabilizer for Magnetocentrifugal
Winds?}
\label{discussion}

The magnetocentrifugal wind in our simulation appears stable in 3D.
There is no evidence for rapid growth of instabilities that would
lead to flow disruption, despite the fact that the magnetic field
in the wind is toroidally dominated all the way to the launching
surface. In particular, there is not hint of exponential growth of
the kink ($m=1$) mode, even though our perturbation at the base
of the wind is designed to maximize the $m=1$ component. The same
conclusion appears to hold for the more lightly loaded winds that
we have done as well (see also \citealt{klb03b}), although
in these cases we are unable to run the simulations for as long.

The most likely reason for the stability is that, in our simulations,
the perturbed magnetocentrifugal part of the wind encloses a (light)
fast-moving outflow near the axis with a poloidally dominated magnetic
field. Mathematically, the fast axial flow is used to provide a
clean inner boundary to the outer part of wind driven through the
magnetocentrifugal
mechanism, which fails close to the axis because of unfavorable field
line inclinations. Physically, it may represent a flow driven
non-magnetocentrifugally along open field lines anchored on young
stars, perhaps by nonlinear Alfv\'en waves generated through magnetic
footpoint motions (e.g., \citealt{si05}). A magnetically
dominated stellar outflow is also seen in the simulations of
\citet{u06} for disk-magnetosphere interaction in the ``propeller''
regime. One may attempt to test the supposition by removing the
poloidally dominated fast outflow in the axial region.
However, if we were to do this, the field line originally at
the interface between the inner flow and outer wind would bend inwards,
pulling the field lines right outside of it to a more vertical
position. The magnetocentrifugal mechanism would shut off for
these field lines, leaving them loaded with little material. A
more tenuous outflow may still be possible along these unfavorably
inclined field lines, driven for example by Alfv\'en waves. We
would then be back to essentially the original
configuration: namely, a lighter and perhaps faster flow dominated
by poloidal magnetic field enclosed by a more heavily loaded wind
that becomes increasingly toroidally dominated at large distances.
Our simulations suggest that the same lightly loaded, nearly vertical
field lines that force
the open field lines further out on the disk to bend by more than
30$^\circ$ away from the axis to make the magnetocentrifugal
wind-launching possible in the first place may stabilize the
launched wind at the same time. In other words, if a wind is
driven magnetocentrifugally, its stability may be guaranteed by the
built-in stabilizer. This two-component structure is intrinsic to
the X-wind theory, where the stabilizer is envisioned as the open
stellar field \citep{os95}. The same stabilizing mechanism
should work equally well, if not better, in the disk-wind picture
where, in addition to the stellar field, lightly loaded field
lines on the inner part of the disk can also contribute to wind
stabilization, especially if there is magnetic flux accumulation
due to disk accretion.

Our simulations are limited to the region of acceleration and early
propagation of magnetocentrifugal winds, where they appear stable.
Whether such winds can stably propagate to much larger distances
remains to be determined. Kink instabilities are seen in some
simulations of MHD jet propagation (e.g., \citealt{nm04}).
The adopted jets are different from the magnetocentrifugally
driven jet in our picture, which is simply the densest part of
a space-filling wind that also includes an inner, axial region
dominated by a poloidal magnetic field and an outer, more tenuous
wide-angle component \citep{shu95}.
We have treated the disk as a fixed boundary. Allowing
the disk to evolve in response to angular momentum removal by
the wind may lead to instability in the coupled disk-wind
system (\citealt{l94}; see, however, \citealt{k04}). We plan
to address this important issue in the future through numerical
experiments.

\acknowledgements
This work was supported in part by NASA grants NAG 5-7007, 5-9180,
5-12102 and NNG05GJ49G, and NSF grants AST 00-93091 and 0307368.


\end{document}